\documentclass[twocolumn,pre,amsmath,amssymb,floatfix,superscriptaddress,showpacs]{revtex4}
\usepackage{graphicx}

\newcommand{\ie}{\emph{i.e.}}

\begin{document}

\title{Periodic States, Local Effects and Coexistence in the BML Traffic Jam Model}
\author{Nicholas J. Linesch}
\affiliation{Center for Computational Science and Engineering, University of California, Davis, CA 95616}
\affiliation{Department of Mathematics, University of California, Davis, CA 95616}
\author{Raissa M. D'Souza}
\affiliation{Center for Computational Science and Engineering, University of California, Davis, CA 95616}
\affiliation{Department of Mechanical and Aeronautical Engineering, University of California, Davis, CA 95616}

\begin{abstract}
The Biham-Middleton-Levine model (BML) is simple lattice model of traffic flow, self-organization and jamming.  Rather than a sharp phase transition between free-flow and jammed, it was recently shown that there is a region where stable intermediate states exist, with details dependent on the aspect ratio of the underlying lattice.  
Here we investigate square aspect ratios, focusing on the region where random, disordered intermediate (DI) states and conventional global jam (GJ) states coexist, and show that DI states dominate for some densities and timescales.  Moreover, we show that periodic intermediate (PI) states can also coexist. PI states converge to periodic limit cycles with short recurrence times and were previously conjectured to arise from idiosyncrasies of relatively prime aspect ratios.  The observed coexistence of DI, PI and GJ states shows that global parameters, density together with aspect ratio, are not sufficient to determine the full jamming outcome. We investigate additional features that lead towards jamming and show that a strategic perturbation of a few selected bits can change the nature of the flow, nucleating a global jam. 
\end{abstract}

\date{\today}

\maketitle

\section{Overview: BML and results}
\begin{figure*}[bt]
\vbox{
\hfill{(a)
{\resizebox{1.8in}{!}{\includegraphics{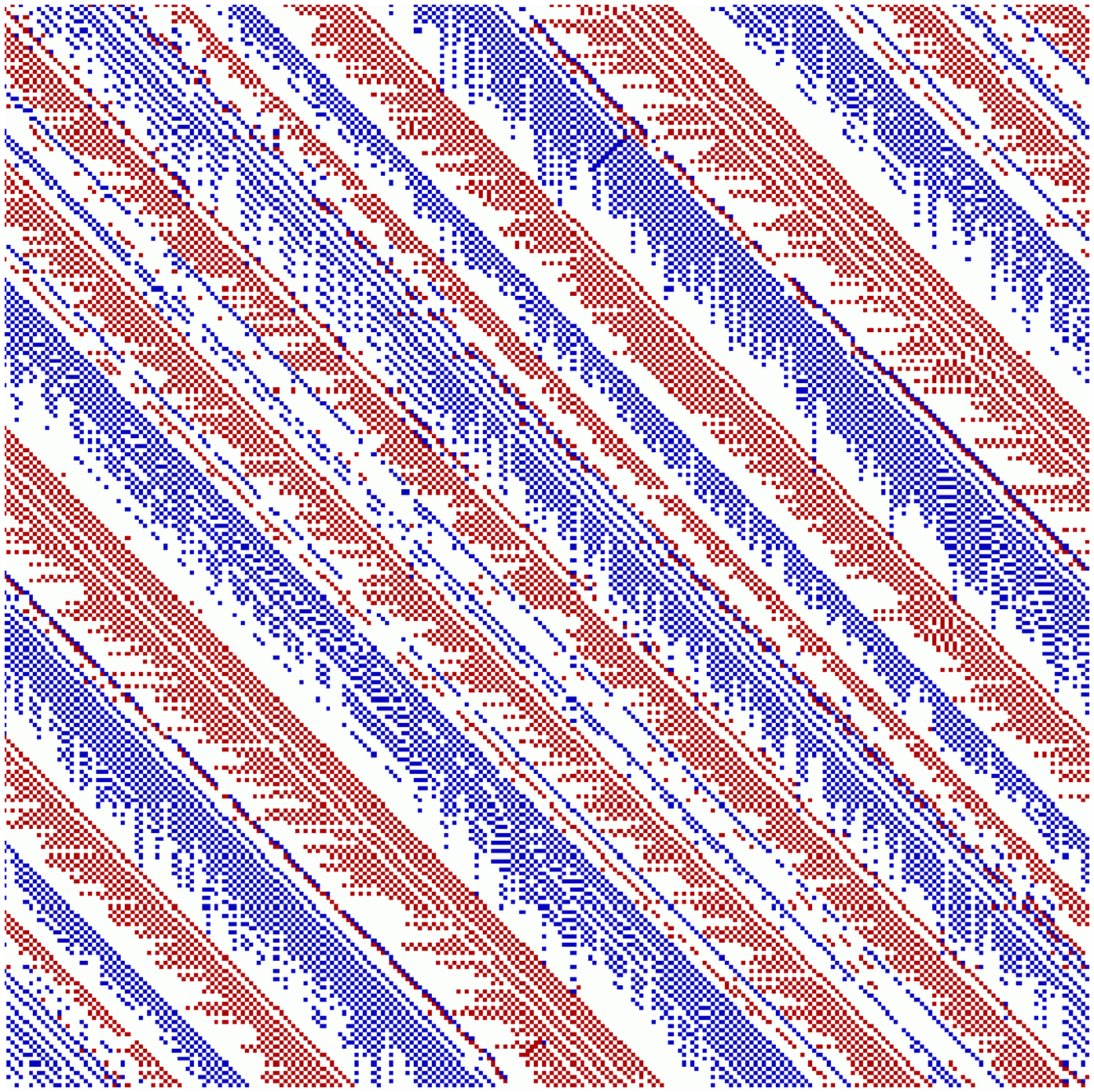}}}}\hfill\hfill
{(b)
{\resizebox{1.8in}{!}{\includegraphics{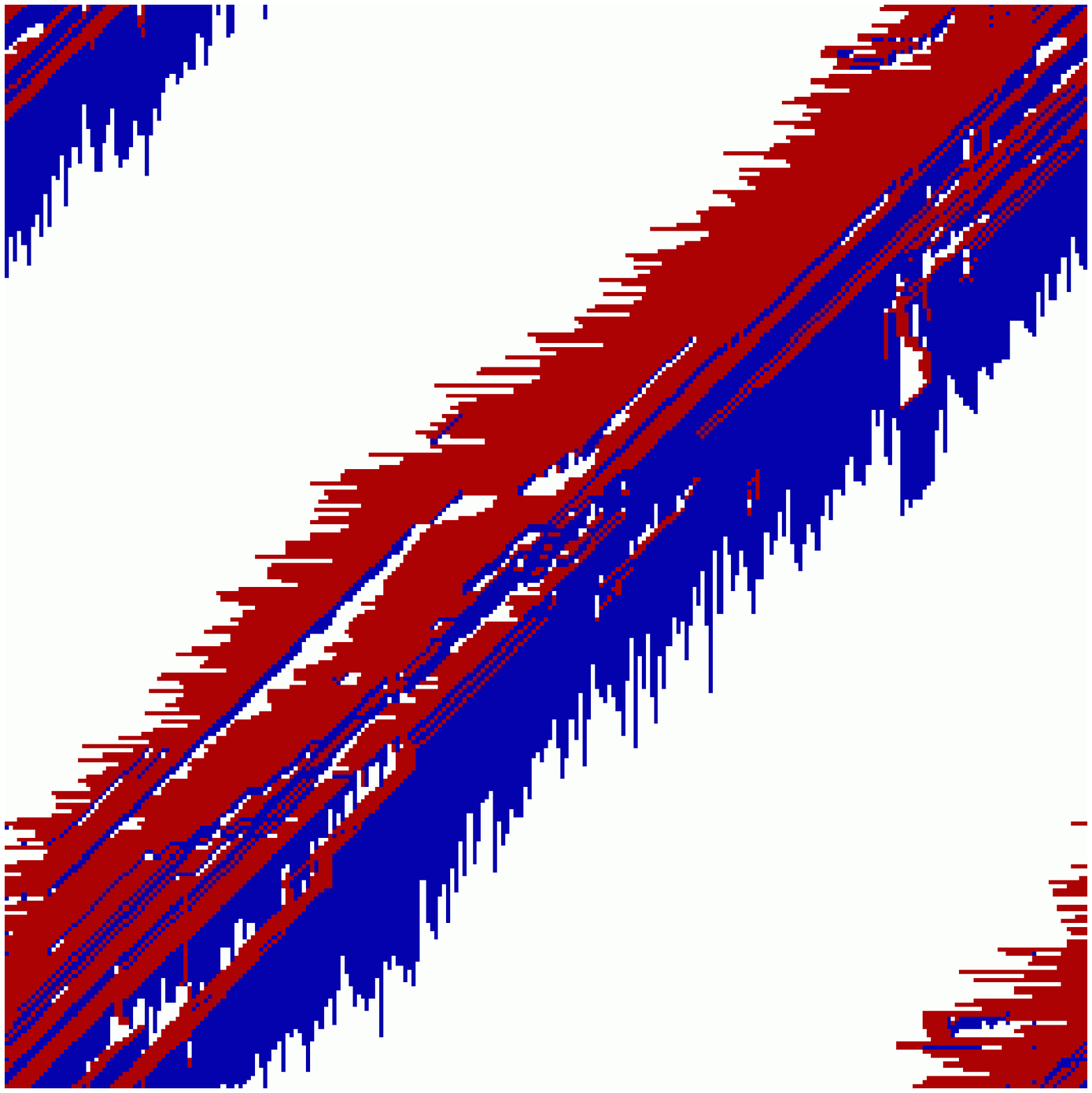}}}}\hfill\hfill
{(c)
{\resizebox{2.0in}{!}{\includegraphics{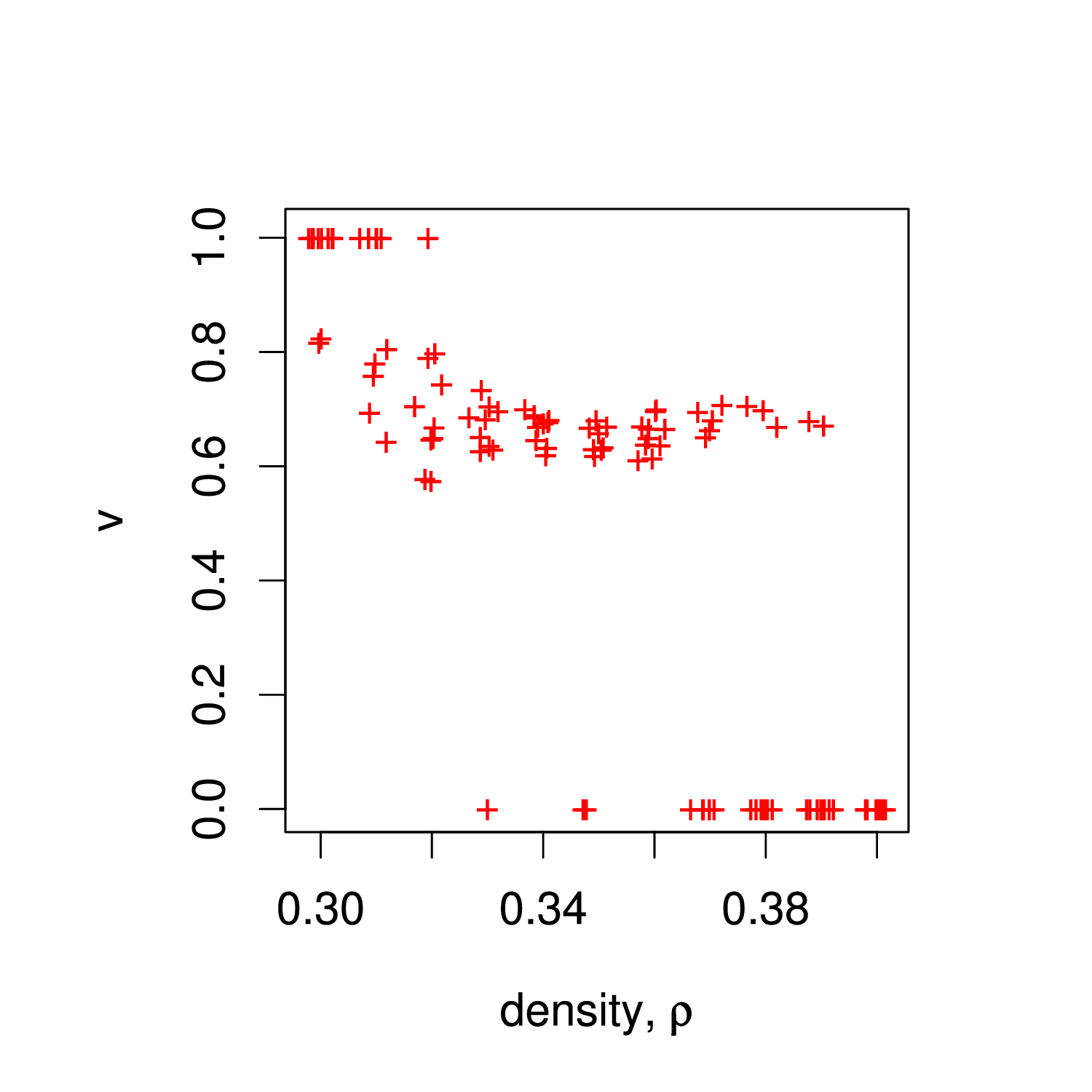}}}}\hfill\hfill
}
\caption{Typical configurations for the BML model on an $L\times L$
lattice, with $L=256$. (a) The free-flow (FF) phase reached by any random
initial configuration with small density, $\rho$. Self-organized
bands of separated east- and north-bound cars allow all particles to
advance each update, thus $v=1$. (b) The global jam (GJ) phase reached by
any random initial configuration for larger $\rho$. Here all cars
are immobilized, thus $v=0$. The conventional belief is that there
is a first-order phase transition between these two behaviors as a
function of $\rho$, with critical density $\rho_c\approx 0.35$. (c) Numerical results show otherwise.  Each point represents one individual random realization run until convergence.  FF ($v=1$) and GJ ($v=0$) states do occur, but less frequently than intermediate states.}
\label{fig:bml}
\end{figure*}

Transport phenomena plays an underlying role in a broad range of physical systems: traffic flow on highways~\cite{HelbingNagel,NagaRev2002,SchadRev2002,ChowSanSchad-Rev2000}, congestion of packets on the Internet~\cite{Jacobson95}, flow of nutrients through the body~\cite{HaffNature2000}, formation and flow of river networks~\cite{BanavarPRL97,BanavarNature99}, etc. 
These all rely on transportation and flow of physical substances.   Self-organized patterns in the flow, such as vortices and spiral waves, occur frequently. Moreover, the flow often jams (i.e., comes to a complete halt) abruptly, in response to just a small change in an external control parameter.  Simple models from statistical physics can capture aspects of self-organization and the abrupt onset of jamming, in particular the Biham, Middleton, and Levine model (BML)~\cite{BML92}. BML is a simple cellular automata model of two-dimensional traffic flow, modeling gridlock between east-bound and north-bound  ``cars". The standard understanding is that the BML model undergoes a first-order phase transition as a function of traffic density $\rho$, from free-flow (FF) traffic (all cars move at all times with velocity $v=1$) to a global jam (GJ) of traffic (no car ever moves, so $v=0$). Figures~\ref{fig:bml}~(a) and (b) show typical FF and GJ configurations. The dynamics leading to their formation can be seen at~\cite{RD-bml-web}.
BML has become a theoretical underpinning for traffic modeling, with  
hundreds of citations in the scientific literature referencing BML and its first-order phase transition. For comprehensive reviews see Refs.~\cite{NagaRev2002,SchadRev2002,ChowSanSchad-Rev2000}.

\begin{figure}[tb]
\begin{center}
(a)~\fbox{\resizebox{2.8in}{!}{\includegraphics{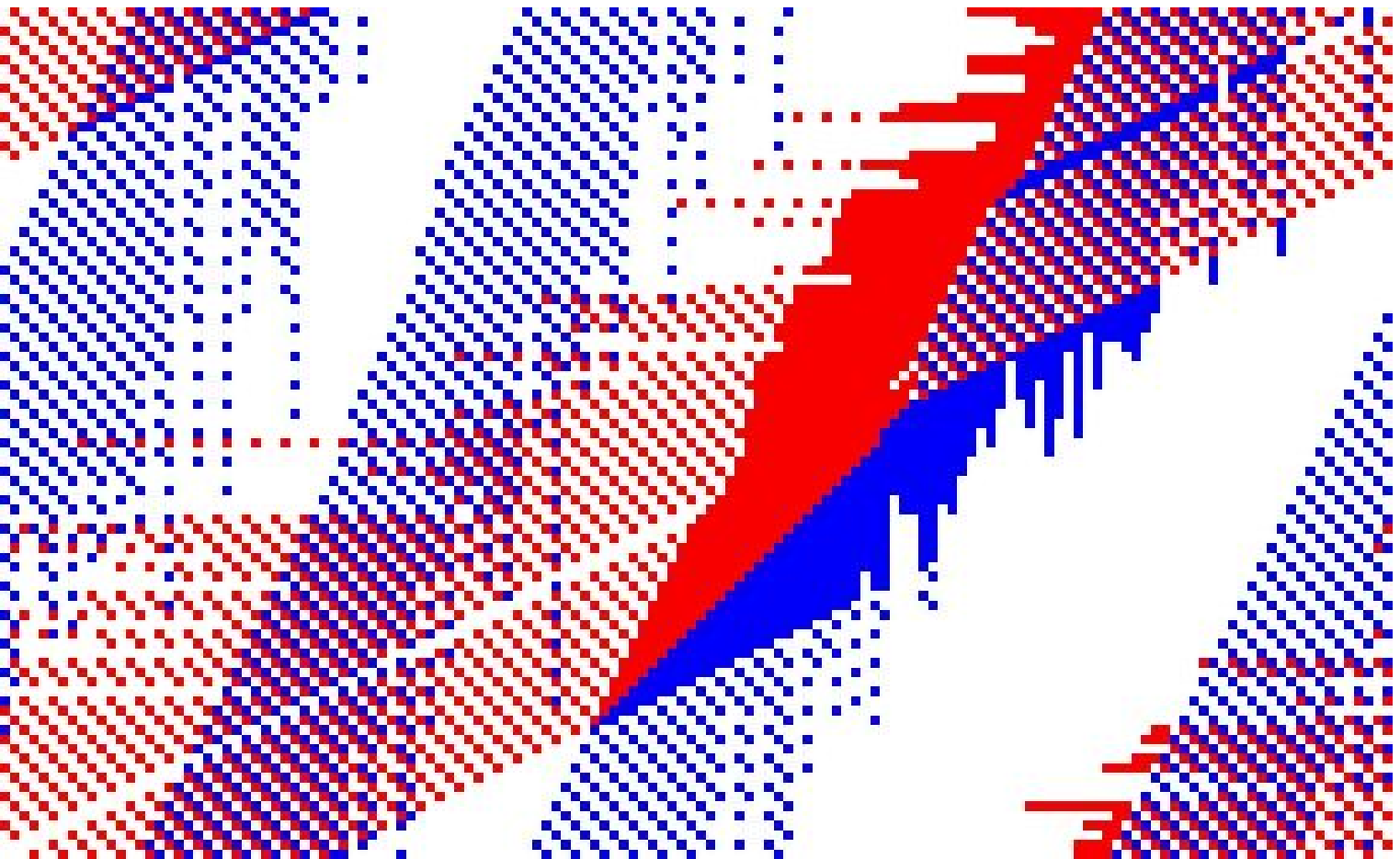}}}\hfill\hfill
\ \vspace{0.2in}\\
(b){\resizebox{2.3in}{!}{\includegraphics{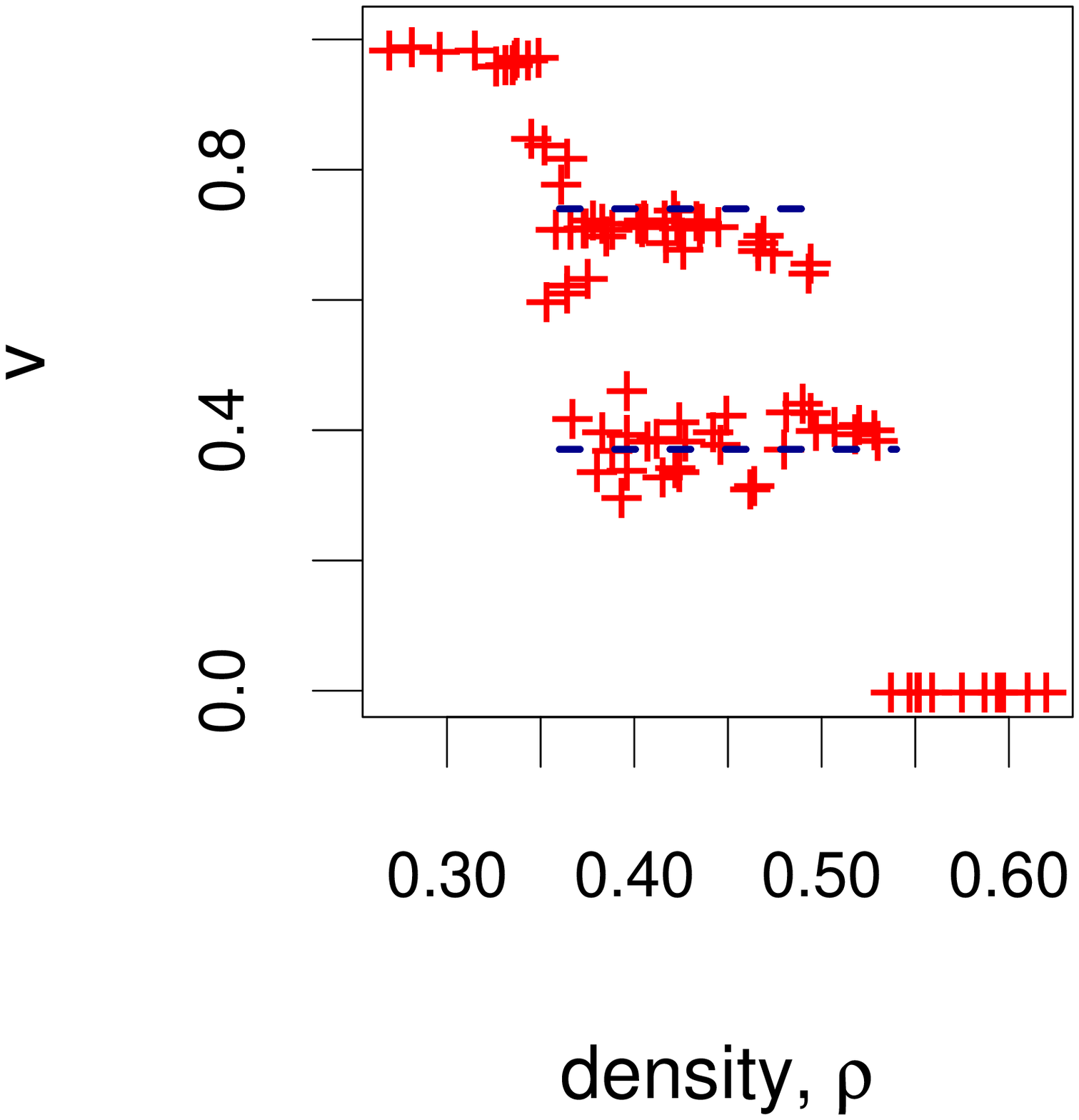}}}\hfill\hfill
\caption{(a) Typical PI state on a Fibonacci lattice for a realization with $L=144$ and $L'=89$. 
(b) Experimental results for $L=89$ and $L'=55$.  Each point represents one individual realization. For densities between $\rho \approx 0.35$ and $\rho \approx 0.5$, 
each realization converges to a PI state.}
\label{fig:fibo}
\end{center}
\end{figure}

It is interesting to note the behavior of the BML model is aspect-ratio dependent and recently a larger family of behaviors was discovered~\cite{RD-BML05}. BML does not necessarily (or even typically, as shown below) exhibit a sharp phase transition from FF to GJ, but instead has a range of intermediate states with regions of FF intersecting at jammed wavefronts.  The geometry and velocities of the intermediate states depend upon the aspect ratio of the underlying lattice (\ie, the ratio of the number of columns, $L$, to the number of rows, $L'$).  When $L$ and $L'$ are two subsequent Fibonacci numbers (referred to herein as ``Fibonacci lattices"), the intermediate states each converge to a periodic limit cycle as was shown in~\cite{RD-BML05}.  The exact microscopic configuration recurs every $\tau$ timesteps, with small $\tau$ on the order of the number of particles in the system.  These periodic intermediate (PI) states have a highly regular, crisp geometric structure, as illustrated in Fig.~\ref{fig:fibo}~(a), with bands of free flowing traffic intersecting at jammed wave fronts that propagate smoothly through the space. (System 1 of Ref.~\cite{RD-bml-web} illustrates this behavior.) Furthermore, as shown in~\cite{RD-BML06} and reproduced here in Fig.~\ref{fig:fibo}~(b), hundreds of runs were simulated for Fibonacci lattices with densities $\rho$ between $1/3$ and $1/2$.  Each one converged to a periodic limit cycle with either $v\approx 0.70$ or $v\approx 0.36$ (see~\cite{RD-BML05} for a derivation of these velocities). The conventional behaviors (FF or GJ) were never once observed and do not seem to be in the accessible phase space for these densities.  As $\rho$ increases there are sharp transitions from the FF phase to the PI to the GJ, with no observed coexistence of these phases.  

\begin{figure}[tb]
\begin{center}
\fbox{\resizebox{1.8in}{!}{\includegraphics{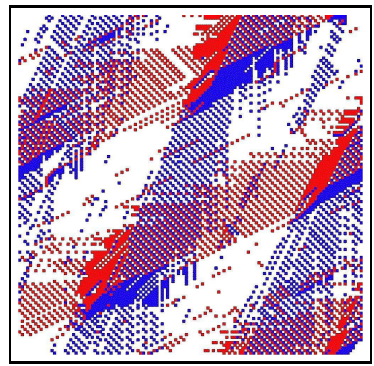}}}
\caption{A typical DI state with $L=L'=128$ and $\rho=0.36$. Chains of particles are interspersed randomly throughout, in contrast to the PI state shown in Fig.~\ref{fig:fibo}~(a).}
\label{fig:DI}
\end{center}
\end{figure}

On square aspect ratios (\ie, $L=L'$, the configuration typically studied in the literature) intermediate states are also observed.  Their typical geometric structure, illustrated by Fig.~\ref{fig:DI},  is qualitatively similar to Fig.~\ref{fig:fibo}~(a), but in addition disordered chains of particles, interspersed at random, move throughout the space and preclude such states from being on periodic limit cycles. (System 3 of~\cite{RD-bml-web} illustrates the dynamics.)  Thus we distinguish Disordered Intermediate (DI) from Periodic Intermediate (PI) states.  Unlike the PI states which have converged to a fixed limiting behavior, we cannot rule out the possibility that the DI states are metastable long-lived transients.  Here we explore convergence properties for intermediate timescales 
and find DI states are extremely common (over 60\% of random realizations with $\rho=0.36$ converge to DI states). In addition, we quantify how stable the DI states are to perturbation.  Here too, the results are consistent with DI states comprising over 60\% of the ensemble of realizations.  

More surprisingly, as reported herein, we discover PI states on square lattices. PI states on square aspect ratios are rare (we observed only two in a sample of 361 random realizations). Yet, their existence shows that stable intermediate states, which have converged to their ultimate limiting behavior,
can exist in the conventional BML model and are not just an idiosyncrasy of relatively prime aspect ratios as previous evidence had suggested~\cite{RD-BML05,RD-bml-weborig}. Furthermore, as  DI, PI and GJ phases can all be observed at the same value of density (in contrast to Fibonacci lattices where there is no observed phase coexistence), we conclude that density together with aspect ratio influence the jamming outcome, but are not ``control parameters", sufficient to fully predict it.   {\it We show that local effects play a fundamental role in determining the outcome}.  A strategic perturbation of just a few bits (which cannot, by definition, alter global properties such as density) can make a realization flip from the DI to the GJ state.

\section{Implementation procedure}\label{implementation}
The BML model consists of two species of ``cars" moving on a two-dimensional square lattice with periodic boundary conditions. ``Red" cars want to move east-ward. ``Blue" cars want to move north-ward. And they alternate attempts to do so. First all the red cars in synchrony attempt to advance one lattice site to the east. Any car succeeds so long as the site it wants to occupy is currently empty (no red or blue car is already occupying it). Then all the blue cars in synchrony attempt the corresponding advance north-ward. Cars that advance are said to have individual velocity $v_i = 1$. Stationary (\ie, ``blocked") cars have $v_i = 0$. 
We can think of the BML lattice as having a traffic light at each site, with all lights synchronously timed to alternate between east-ward and north-ward flow. The dynamics is fully deterministic. The only randomness is in the initial condition, when an empty lattice is populated with cars. 

\begin{figure}[b]
\begin{center}
{\resizebox{2.5in}{!}{\includegraphics{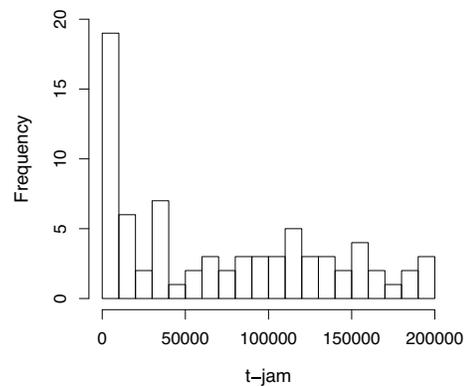}}}
\caption{Histogram of jamming times for an ensemble of 180 random realizations with $L=L'=128$ and $\rho=0.36$.}
\label{fig:tjam}
\end{center}
\end{figure}

We implement the BML model using a square aspect ratio with $L=L'=128$. Each lattice site is initialized independently at random and filled with a red car with probability $\rho/2$, and then a blue car with probability $\rho/2$, with overall $\rho=0.36$.   
This random realization is then run until it jams or until reaching $\tau_c = 2\times 10^5$ time-steps. (Note each unit time-step consists of both a red sub-step and a blue sub-step.)  In~\cite{RD-BML05} the median convergence time for systems of size $L=128$ was determined to be approximately $5\times 10^5$.  We use the shorter time here in order to sample more configurations and thus specifically explore convergence properties for $t < 2\times 10^5$.  
A histogram of jamming times for an ensemble of 180 random realizations in shown in Fig.~\ref{fig:tjam}. 40\% of the systems jam in time $t< \tau_c$, with approximately 1/3 of those jamming in time $t<20,000$, and over 4/5's jamming in time $t<150,000$.  While we expect that many DI states will jam eventually, it was possible for us to simulate a few configurations for orders of magnitude longer than $\tau_c$ and observe the DI phase persisting.

At any particular time-step the overall system velocity $v_t$ is the average over the individual velocities of all $N$ cars present: $v_t = \sum_{i=1}^N v_i(t)/N$.  For a GJ, $v_t=0$. For all other systems observed once $t=2\times 10^5$, $v_t$ is quite stable, with only small fluctuations. 
To average over the fluctuations, the value of velocity reported herein, $v$, is the average of $v_t$ over a window of length $\Delta t = 100$, measured from $t=200,000$ to $t=200,100$. 

All simulations were implemented in Python~\cite{python}. The numerical results presented in Fig.~\ref{fig:ensemble} and Fig.~\ref{fig:flip} alone required over a month of non-stop running time, using an 8-processor, dual-core, 2.4GHz computer.  Python allows us to visualize the system as it is being simulated. Although this slows down the simulations, visualization is the key component for discovering the new phenomena reported herein, in particular the existence of PI states.

\section{Results}\label{results}
\subsection{Accessibility of intermediate states}

\begin{figure}[t]
\begin{center}
\resizebox{2.8in}{!}{\includegraphics{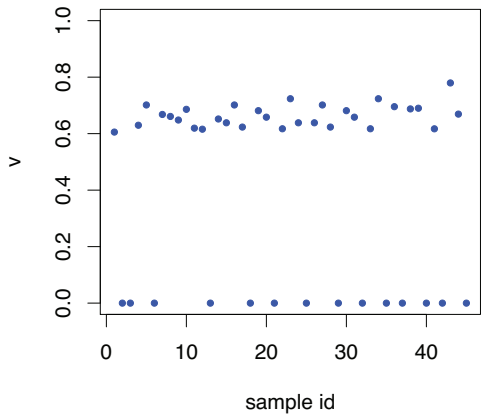}}
\caption{Experimental results for the $45$ realizations with $\rho=0.36$ on an $L=L'=128$ lattice.  Each point represents one individual realization started from a random initial condition and run until time $\tau_c$.  Only 14 realizations converge to the expected GJ state.  The remaining $31$ converge to the DI state.}
\label{fig:ensemble}
\end{center}
\end{figure}

We conduct a detailed study starting with 45 random initial conditions. 14 of these converge to the jammed state ($v=0$).  The remaining $31$ converge to the DI state (with $v \approx 2/3$).  Figure~\ref{fig:ensemble} is a plot of the exact converged velocity for each sample.  Thus for $L=L'=128$ and densities $\rho\approx 0.36$, more than than $68\%$ of the random initial conditions converge to the DI state. Less than $32\%$ of them converge to the expected GJ behavior of $v=0$. 

\subsection{Robustness of DI states}\label{perturb}
We wish to understand how robust the intermediate states are to perturbation.  If we flip the identity of some of the cars (\ie, from red-to-blue and blue-to-red), how likely is the perturbed system to then jam?  We consider the subset composed of the $31$ DI realizations discussed above.  
For each realization, we choose some fraction of the particles uniformly at random (ensuring equal numbers of red and blue), and exchange their color.  Then we run the perturbed system for a second $\tau_c$ timesteps and measure the new velocity.  

\begin{figure}[t]
\begin{center}
\resizebox{2.7in}{!}{\includegraphics{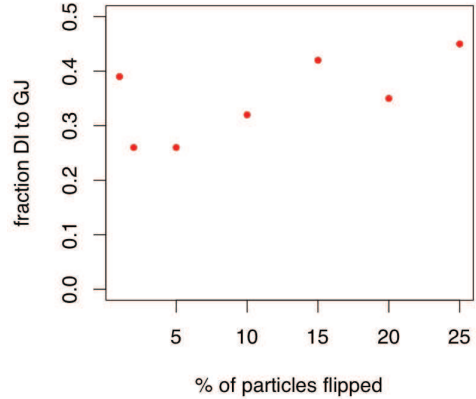}}
\caption{The fraction of realizations which flip from DI to GJ as a function of the extent of the perturbation.}
\label{fig:flip}
\end{center}
\end{figure}

Figure~\ref{fig:flip} is a plot of the fraction of intermediate states which changed from DI to GJ as a function of the percentage of the particles whose identities were flipped.  Between 25\% and 45\% of the realizations flip from DI to GJ when perturbed, but the majority remain in the DI phase.  This is consistent with the findings shown in Fig.~\ref{fig:ensemble}, that DI states comprise approximately 60\% of the phase space. 
Note, in Fig.~\ref{fig:flip}, random perturbations as large as 25\% of the particles are considered and there is a weak trend indicating a system is more likely to reach a GJ with increased perturbation.
But, more surprisingly, as will be presented in Sec.~\ref{sec:jam}, a tiny {\em strategic} perturbation can take a realization from the DI to the GJ phase. 

\subsection{PI states and phase stability}
\begin{figure}[t]
\begin{center}
\fbox{\resizebox{1.8in}{!}{\includegraphics{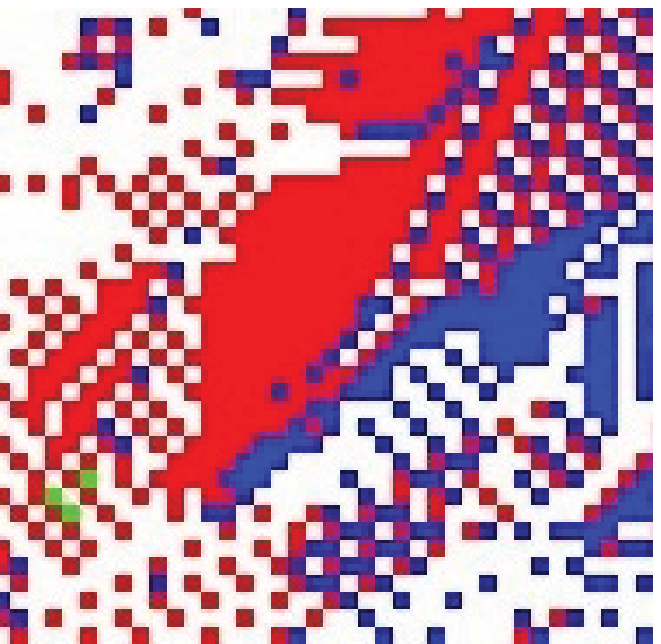}}}
\caption{(Figure at end of manuscript.) A typical PI state on a square lattice for a realization with $L=L'=128$ and $\rho=0.36$.  System 2 of~\cite{RD-bml-web} illustrates the dynamics. }
\label{fig:sq-per}
\end{center}
\end{figure}

Two of the DI states, when perturbed slightly (less than 5\%), 
actually went from DI to PI within the $\tau_c$ time-steps simulated. Both of these PI states have $v\approx 0.5$ and period, $\tau \approx 25,000$ timesteps (about four times the number of particles in the system).  A snapshot of one of the PI realizations is shown in Fig.~\ref{fig:sq-per}. (System 2 of Ref~\cite{RD-bml-web} is a movie of the dynamics.)  Our total sample space includes 361 realizations (\ie, the original ensemble of 45 plus all the perturbed realizations). Of these 361, only two went to a PI state, thus PI states are rare. Furthermore, the PI states observed are very fragile. Flipping just $1\%$ of the particles takes the system to the DI phase.  

It is interesting to compare this behavior with that for Fibonacci lattices. For Fibonacci lattices, individual PI states are also fragile. If just one particle of a particular PI realization is perturbed, that realization jumps to a different PI state (\ie, a different periodic limit cycle).  But note, it remains in the PI phase.  Thus, an individual PI state is fragile, yet the PI phase is extremely robust.  For square lattices, each PI state is fragile and moreover the PI phase is fragile. Note, for both Fibonacci and square lattices the GJ phase is extremely stable to perturbation.  Any limited perturbation affects cars located primarily in the interior of the jammed region which remain jammed.  The standard BML model is known to show some dependence on systems size (the onset of jamming occurs at a density which decreases slightly with increasing system size $L$~\cite{BML92,RD-BML06}).  An open question is to examine the change in frequency of occurrence of DI and PI states with increasing system size. 

\section{Local properties trigger jamming} 
\label{sec:jam}

Jamming and transport play an underlying role in a wide array of fundamental processes, hence identifying properties that enhance, delay or trigger jamming is an important problem.  
As shown in Sec.~\ref{results}, systems with $L/L'=1$ and the same exact $\rho$ may converge to GJ, to DI or to PI states. Thus, $\rho$, together with $L/L'$, predict some,  but not all,  properties of the jamming outcome.  
Our aim is to identify the additional factors that influence jamming -- to find properties that differentiate between realizations which, for the same $\rho$ and $L/L'=1$, converge to GJ from those which converge to DI states. We investigate a range of potential factors, amongst them:\\
- The maximum density observed in any one row or column. The assumption is that a highly populated row or column could nucleate a jam.\\
- The standard deviation of the distribution of individual row (and column) densities.  The assumption is that large variance (many high density lines mixed with low density ones) can nucleate a global jam. \\
- The difference between the total number of red- versus blue-cars. The assumption is that a disproportionate representation of one color could nucleate a jam.  Note, each site is occupied initially at random with probability $\rho$, hence the total number of occupied sites is a random variable.

First we focus on the original ensemble of 45 realizations (defined in Sec.~\ref{implementation}).  We found no correlation between the values of any of these properties listed above and the likelihood to converge to a GJ, rather than a DI, state.  Furthermore, we found no correlation between any of these properties and the value of the converged velocity (Fig.~\ref{fig:ensemble} is a plot of the latter).  These macroscopic properties are independent from the likelihood to jam. 
Similar results were found in a previous study on Fibonacci lattices, where there are two types of PI states (high- and low-velocity), and we found that none of the properties above differentiated between them~\cite{RD-unpublished}. 
  
We then focus on comparing individual realizations to their perturbed variants (as defined in Sec.~\ref{perturb}).  In particular we consider realizations that, with just a 1\% perturbation, change from DI to GJ.  We find the properties discussed above do not differentiate between the original and perturbed realizations.  To clarify whether these particular realizations that went from DI to GJ are more susceptible to jamming we tried several different random 1\% tweaks, but most did not nucleate a jam. Rather then being an intrinsic property of the realization, it seems that sometimes we happen to perturb the key particles and nucleate a jam.

\begin{figure}[t]
\begin{center}
\fbox{\resizebox{1.8in}{!}{\includegraphics{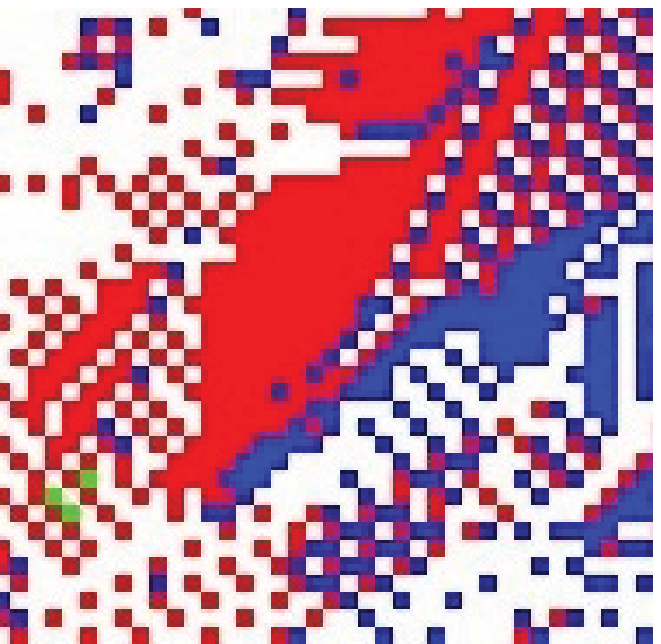}}}
\caption{Closeup of the DI state shown in Fig.~\ref{fig:DI}, with $L=L'=128$ and $\rho=0.36$.  (To view this figure, please download the journal version of this manuscript from http://mae.ucdavis.edu/dsouza/pubs.html -- many difficulties uploading these figures to the arxiv.) Perturbing just three particles (colored green, lower left hand corner), will make the new system converge to GJ. (System 4 of Ref.~\cite{RD-bml-web} is a movie of the dynamics, started from the perturbation.)}
\label{fig:three-flip}
\end{center}
\end{figure}

Diagonal ordering may be of fundamental importance to the BML model.  The self-organization leading to FF involves formation of left-sloping (NW-to-SE) diagonals of the same species (Figure~\ref{fig:bml}~(a) illustrates the end result of this process, with separated left-sloping bands of the different species).  Jams, in contrast, form along right-sloping (NE-to-SW) diagonals.  (Figure~\ref{fig:bml}~(b) illustrates the end result of a global jam).  During the evolution of any realization, from the random initial condition to the final state, we see a competition between these two types of diagonals emerging and interfering with each other.  The left-sloping diagonals lead towards $v=1$, the right-sloping towards $v=0$.  Note, Austin and Benjamini recently completed an analysis of BML based on diagonals~\cite{AusBenj06}.  They showed that, for an $L\times L$ system with total number of particles $N=L/2$ the system always converges to $v=1$.  Of course, $\rho=1/(2L)$ is quite far from the regime of interest ($0.3 < \rho < 0.6$), yet~\cite{AusBenj06} is one of the only rigorous proofs for behavior of the BML model.  

For both DI and PI states, there are regions of left-sloping bands of FF intersecting at right-sloping jammed interfaces. By perturbing just a few particles in nearby left-sloping bands, we can nucleate a jam.  Figure~\ref{fig:three-flip} shows an example realization. We start from the same system shown in Fig.~\ref{fig:DI} and then exchange the velocity of just {\it three} particles along ordered left-sloping diagonals, which makes the realization flip from the DI to GJ phase. (System 4 of Ref.~\cite{RD-bml-web} is a movie of the process.) Here, perturbing all three particles is necessary in order to induce the GJ phase, however we did observe another realization where changing just {\it one} particle made the realization flip from DI to GJ.  (Note that a 1\% perturbation of a realization with $L=L'=128$ and $\rho=0.36$ amounts to perturbing approximately $60$ particles.  Perturbing three particles represents just a $0.01\%$ change.)  
We believe that pinpointing exactly how to quantify the change induced by the local perturbation is a difficult challenge, and leave this as an open question.

\section{Discussion}
The phase space of the BML model is much richer than previously expected, displaying a range of intermediate behaviors. Here we have shown that car density, $\rho$, together with aspect ratio influence the jamming outcome, but are not necessarily sufficient to fully determine it.  For Fibonacci lattices, three distinct phase regimes are observed as a function of increasing $\rho$: 1) FF, 2) PI and 3) GJ.   Thus the value of $\rho$ is predictive of the overall phase a realization will converge to (FF, PI or GJ).  For square aspect ratios the regimes observed are: 1) FF, 2) coexistence of PI and DI, and GJ and 3) GJ.  The coexistence of PI, DI and GJ states for the same value of $\rho$ means $\rho$ is not sufficient to determine the phase a realization will converge to.   Local properties also seem to play an important role, and we present evidence that diagonal ordering may be an additional, fundamental consideration. 

The intermediate states on the square aspect ratio comprise a significant fraction of the accessible phase space. For $\rho=0.36$, over 60\% of the states observed are in the DI phase for the timescales we measured.  
Furthermore, results from perturbing DI states also are consistent with such phases comprising over 60\% of the phase space.  

Our current study deals with perturbing a realization after it has converged to the DI phase. Far more powerful (and difficult) would be to understand what to alter in the random initial condition to control the phase that realization ultimately converges to. 
Another outstanding challenge is to pinpoint exactly what was achieved with our perturbation.

A practical implication of this work that may be worth considering involves the onset of jamming. For square aspect ratios GJ states first appear at $\rho \sim 1/3$ (as shown in Fig.~\ref{fig:bml}~(c)), whereas for Fibonacci lattices the GJ phase is not observed until $\rho > 1/2$. Thus, by changing the underlying lattice aspect (from square to Fibonacci), one can considerably delay the onset of when complete jamming is first observed, and instead observe high-throughput flow in a regime where previously it was believed the system would fully jam.

\section{Acknowledgments}  
We thank Benny Brown for Python programming tips, and both he and Soumen Roy for useful discussions.  We are grateful to Dov Levine for initially suggesting we investigate stability of intermediate states and to the anonymous referees for many useful comments.

\end{document}